\documentclass[prd,superscriptaddress,nofootinbib,amsmath,amssymb,aps,11pt]{revtex4}

\usepackage{bm}
\usepackage{amsfonts}
\usepackage{latexsym}
\usepackage[latin1]{inputenc}
\usepackage{graphicx}
\usepackage{amsmath}
\usepackage{palatino}
\usepackage{mathpazo}
\usepackage{booktabs}
\usepackage{dcolumn}

\def\jnl@style{\it}
\def\aaref@jnl#1{{\jnl@style#1}}

\def\aaref@jnl#1{{\jnl@style#1}}

\def\aj{\aaref@jnl{AJ}}                   
\def\apj{\aaref@jnl{ApJ}}                 
\def\apjl{\aaref@jnl{ApJ}}                
\def\apjs{\aaref@jnl{ApJS}}               
\def\apss{\aaref@jnl{Ap\&SS}}             
\def\aap{\aaref@jnl{A\&A}}                
\def\aapr{\aaref@jnl{A\&A~Rev.}}          
\def\aaps{\aaref@jnl{A\&AS}}              
\def\mnras{\aaref@jnl{Mon.~Not.~Roy.~Astron.~Soc.}}             
\def\prd{\aaref@jnl{Phys.~Rev.~D}}        
\def\prc{\aaref@jnl{Phys.~Rev.~C}}  
\def\prl{\aaref@jnl{Phys.~Rev.~Lett.}}    
\def\qjras{\aaref@jnl{QJRAS}}             
\def\skytel{\aaref@jnl{S\&T}}             
\def\ssr{\aaref@jnl{Space~Sci.~Rev.}}     
\def\zap{\aaref@jnl{ZAp}}                 
\def\nat{\aaref@jnl{Nature}}              
\def\aplett{\aaref@jnl{Astrophys.~Lett.}} 
\def\apspr{\aaref@jnl{Astrophys.~Space~Phys.~Res.}} 
\def\physrep{\aaref@jnl{Phys.~Rep.}}      
\def\physscr{\aaref@jnl{Phys.~Scr}}       
\def\commat{\aaref@jnl{Comm.~Math.~Phys.}}              
\def\science{\aaref@jnl{Science}}               
\def\cqg{\aaref@jnl{Classical Quant.~Grav.}}            
\def\jpcs{\aaref@jnl{JPCS}}                                     
\def\ijmpd{\aaref@jnl{Int.~J.~Mod.~Phys.~D}}                    
\def\grg{\aaref@jnl{Gen.~Relat.~Gravit.}}               
\def\rpp{\aaref@jnl{Rep.~Prog.~Phys.}}          
\def\npa{\aaref@jnl{Nucl.~Phys.~A}}        
\def\lrr{\aaref@jnl{Living Rev.~Rel.}}                   
\def\jcap{\aaref@jnl{J.~Cosmology Astropart.~Phys.}}    
\def\rmp{\aaref@jnl{Rev.~Mod.~Phys.}}   

\allowdisplaybreaks[1]

\begin{document}

\title{Stability of topological neutron stars }

\author{Daniela D. Doneva}
\email{daniela.doneva@uni-tuebingen.de}
\affiliation{Theoretical Astrophysics, Eberhard Karls University of T\"ubingen, T\"ubingen 72076, Germany}
\affiliation{INRNE - Bulgarian Academy of Sciences, 1784  Sofia, Bulgaria}

\author{Stoytcho S. Yazadjiev}
\email{yazad@phys.uni-sofia.bg}
\affiliation{Theoretical Astrophysics, Eberhard Karls University of T\"ubingen, T\"ubingen 72076, Germany}
\affiliation{Department of Theoretical Physics, Faculty of Physics, Sofia University, Sofia 1164, Bulgaria}
\affiliation{Institute of Mathematics and Informatics, 	Bulgarian Academy of Sciences, 	Acad. G. Bonchev St. 8, Sofia 1113, Bulgaria}

\author{Kostas D. Kokkotas}
\email{kostas.kokkotas@uni-tuebingen.de}
\affiliation{Theoretical Astrophysics, Eberhard Karls University of T\"ubingen, T\"ubingen 72076, Germany}


\begin{abstract}
Tensor-multi-scalar theories (TMST) are among the most natural generalizations of Einstein's theory, they are mathematically self-consistent and free from pathologies. They pass through all the known observations but contrary to standard scalar-tensor theories, TMST offer extremely rich spectrum of solutions and allow for large deviations from general relativity. One of the most interesting objects in these theories are the topological neutron stars. The main goal in the present paper is to study their radial stability since many branches of solutions can exist even for a fixed value of the topological charge. It turns out that only one of these branches is stable for each value of the topological charge while all the rest are unstable. The stable branch is exactly the one that spans from small to large neutron star masses having as well moderate values of the radii in agreement with the observations. As expected, it becomes unstable at the maximum of the mass. The frequencies of the radial modes are examined and it turns out that in most cases the mode frequencies are larger than the general relativistic ones and they increase with the increase of the topological charge. All this shows that the topological neutron stars are viable astrophysical objects that should be explored further in order to determine their observational manifestations.
\end{abstract}

\maketitle

\section{Introduction}
In view of the huge advance in astrophysical observations in the past few years, the aim to probe further the theory of gravity is becoming a realistic one. This mainly concerns the strong field regime that is not well explored observationally. That is why the study of compact objects in alternative theories of gravity is attracting more and more attention recently \cite{Berti2015a,2018ASSL..457..737D}. Among the most natural generalizations of Einstein's theory are classes of alternative theories which posses an additional mediator of the gravitational interaction, namely one or several scalar fields. The simplest choice are the scalar-tensor theories (STT)   which gained significant attention especially with the discovery of the so-called spontaneous scalarization of neutron stars (NS), that is a nonlinear development of a nontrivial scalar field ones a critical threshold for the compactness of the NS is exceeded \cite{Damour1993}. If one takes into account the observational constraints, though, the deviations from GR are very small -- the Brans-Dicke parameter is severely constrained by the weak field observations while the effect of scalarization is strongly limited by the observations of neutron stars in close binary systems \cite{Will:2014kxa,Demorest10,Antoniadis2013a}. 

A way to generalize the classical STT in order to produce compact objects that obey both the weak and the strong field observations, but can produce large deviations from GR, is to allow for the existence of multiple scalar fields. These are the so-called tensor-multi-scalar theories  (TMST) of gravity that are among the most viable alternative gravitational theories --  they can pass all known experimental and observational tests \cite{Damour1992,Horbatsch:2015bua,Yazadjiev:2019oul,Doneva:2019krb}. TMST are also  mathematically self-consistent and free from pathologies contrary to other alternative theories which suffer from such.  Within the framework of these theories the gravitational interaction is mediated not only by the spacetime metric but also by additional scalar fields. These additional $N$ scalar fields $\varphi^a$ take values in a coordinate patch of an abstract $N$-dimensional Riemannian target space ${\cal E}_{N}$ with a metric $\gamma_{ab}(\varphi)$. 

Soliton and mixed fermion-soliton solutions were recently constructed in TMST \cite{Yazadjiev:2019oul,Doneva:2019krb}, including the case of rapid rotation \cite{Collodel:2019uns}. The TMST with $\mathbb{S}^3$ target space were the first alternative theory of gravity where topological neutron stars were found to exist \cite{Doneva:2019ltb}. Scalarized neutron stars with a target space which is a maximally symmetric 3-dimensional space ($\mathbb{S}^3$, $\mathbb{H}^3$ or $\mathbb{R}^3$)   were obtained as well in \cite{PhysRevD.101.104010}. All this shows that the great spectrum of solutions that can be obtained within TMST is controlled by the freedom of choosing the target space ${\cal E}_N$, the target space metric $\gamma_{ab}(\varphi)$, the conformal factor $A(\varphi)$ and the map $\varphi:\text{\it spacetime} \to \text{\it target space}$.   A very important property of the topological neutron stars with $\mathbb{S}^3$ target space and the scalarized neutron stars  with $\mathbb{S}^3$, $\mathbb{H}^3$ or $\mathbb{R}^3$ target space is that, contrary to the standard scalar-tensor theories, the scalar charge is zero. Therefore, no dipole scalar gravitational radiation is present and the binary pulsar constraints do not apply. Similar features were also  observed in other alternative theories  such as the Einstein-dilaton-Gauss-Bonnet theory in decoupling limit~\cite{Pani:2011xm}. 

As we commented, one of the very interesting predictions in TMST is the existence of neutron stars with nonzero topological charge. This offers the possibility for the presence of new phenomena that can be tested via the forthcoming gravitational wave and electromagnetic observations. Instead of having only one branch of solutions like in pure GR, multiple branches of topological NSs exist \cite{Doneva:2019ltb}. This brings the question which are the stable one(s), if any. Certain considerations in \cite{Doneva:2019ltb} has already brought us to the hypothesis that only one of these branches is stable while the rest are unstable. A proper analysis, though, requires to study the dynamical stability of the topological neutron stars against radial linear perturbations that will be the focus of the present paper. 

The radial perturbations of neutron stars in STT were considered in \cite{Sotani2014,Mendes:2018qwo}. In the present paper we will generalize these results to the case of TMST. Other classes of modes in alternative theories of gravity were considered in a number of papers. The fundamental non-radial and pressure modes of NSs in STT and $f(R)$ gravity  were considered in \cite{Sotani04,Staykov2015}, while the axial modes were examined in \cite{Sotani2005,Blazquez-Salcedo:2018tyn,Blazquez-Salcedo:2015ets,Blazquez-Salcedo:2020ibb,AltahaMotahar:2019ekm,AltahaMotahar:2018djk} (see \cite{Blazquez-Salcedo:2018pxo} for a review). In the former case, though, that is also observationally more relevant, most of the studies were done in the Cowling approximation, i.e. when the metric (and in most cases the scalar field) are kept fixed. In the present paper we will consider the full problem of coupled scalar field, metric and fluid radial perturbations.

In Section II we review the basic theory behind TMST and the equilibrium neutron star solutions in these theories while the equations governing the NS perturbations are derived in Section III. The results are presented in Section IV. The paper ends with Conclusions.

\section{Tensor-multi-scalar theories and topological neutron star solutions}
In more precise mathematical language, the gravitational interaction in TMST is mediated  by the spacetime metric $g_{\mu\nu}$ and $N$ scalar fields $\varphi^{a}$  which take value in a coordinate patch of an N-dimensional Riemannian (target) manifold ${\cal E}_{N}$  with (positively definite) metric $\gamma_{ab}(\varphi)$ defined on it \cite{Damour1992,Horbatsch:2015bua}. The Einstein frame action of the  TMST of gravity is the following 
 \begin{eqnarray}\label{Action}
  S=&& \frac{1}{16\pi G_{*}}\int d^4\sqrt{-g}\left[R - 2g^{\mu\nu}\gamma_{ab}(\varphi)\nabla_{\mu}\varphi^{a}\nabla_{\nu}\varphi^{b} - 4V(\varphi)\right]  \nonumber \\
  &&+ S_{matter}(A^{2}(\varphi) g_{\mu\nu}, \Psi_{matter}),
  \end{eqnarray}
  where $G_{*}$ is the bare gravitational constant, $\nabla_{\mu}$ and $R$ are the covariant derivative  and the Ricci scalar curvature with respect to  the Einstein frame metric $g_{\mu\nu}$, and $V(\varphi)\ge 0$ is the potential of the scalar fields. In order for the weak equivalence principle to be satisfied the matter fields, denoted collectively by $\Psi_{matter}$, are coupled only to the physical Jordan frame metric ${\tilde g}_{\mu\nu}= A^2(\varphi) g_{\mu\nu}$ where  $A^2(\varphi)$ is the conformal factor relating the Einstein and the Jordan frame metrics, and which, together with $\gamma_{ab}(\varphi)$ and $V(\varphi)$, specifies the TMST.  From a more global point of view $\varphi^a$ 
  define a map $\varphi : \text{\it spacetime} \to \text{\it target space}$ and the scalar fields kinetic term in the action above is just the pull-back  of the line element of the target space.  
  
The  Einstein frame field equations yielded by the action (\ref{Action}) read  
  \begin{eqnarray}\label{FE}
  &&R_{\mu\nu}= 2\gamma_{ab}(\varphi) \nabla_{\mu}\varphi^a\nabla_{\nu}\varphi^b + 2V(\varphi)g_{\mu\nu} + 8\pi G_{*} \left(T_{\mu\nu} - \frac{1}{2}T g_{\mu\nu}\right), \nonumber \\
  &&\nabla_{\mu}\nabla^{\mu}\varphi^a = - \gamma^{a}_{\, bc}(\varphi)g^{\mu\nu}\nabla_{\mu}\varphi^b\nabla_{\nu}\varphi^c 
  + \gamma^{ab}(\varphi) \frac{\partial V(\varphi)}{\partial\varphi^{b}}  \\
  &&\hskip 2.0cm -  4\pi G_{*}\gamma^{ab}(\varphi)\frac{\partial\ln A(\varphi)}{\partial\varphi^{b}}T, \nonumber
  \end{eqnarray}
  with $T_{\mu\nu}$ being  the Einstein frame energy-momentum tensor of matter and $\gamma^{a}_{\, bc}(\varphi)$ being 
  the Christoffel symbols with respect to the target space metric $\gamma_{ab}(\varphi)$. From the field equations and the contracted Bianchi identities we also find the following conservation law for the Einstein frame energy-momentum tensor
  \begin{eqnarray}\label{Bianchi}
  \nabla_{\mu}T^{\mu}_{\nu}= \frac{\partial \ln A(\varphi)}{\partial \varphi^{a}}T\nabla_{\nu}\varphi^a .
  \end{eqnarray}
  
The Einstein frame energy-momentum tensor $T_{\mu\nu}$ and the Jordan frame one ${\tilde T}_{\mu\nu}$
are related via the formula $T_{\mu\nu}=A^{2}(\varphi){\tilde T}_{\mu\nu}$. As usual, in the present paper  the matter content of 
the stars will be described as a perfect fluid. In the case of a perfect fluid the relations between the energy density,
 pressure and 4-velocity in both frames are given by $\varepsilon=A^{4}(\varphi){\tilde \varepsilon}$, $p=A^{4}(\varphi){\tilde p}$ and 
 $u_{\mu}=A^{-1}(\varphi) {\tilde u}_{\mu}$.

In TMST  various compact objects can be constructed \cite{Yazadjiev:2019oul}--\cite{Doneva:2019ltb} where, as we mentioned, the freedom comes not only from the choice of the conformal factor like the standard scalar-tensor theories, but more importantly from the choice of the target space and the metric defined on it, as well as from the choice of the map $\varphi : \text{\it spacetime} \to \text{\it target space}$. Among them are the so-called topological neutron stars \cite{Doneva:2019ltb}.  In addition to the standard characteristics 
of the usual neutron stars, the topological neutron stars are also characterized by a topological charge. They are static spherically symmetric solution associated with a topologically nontrivial map  $\varphi : \text{\it space} \to \text{\it target space}$. 

We will be concentrated on nonrotating neutron stars and thus the spacetime is static and spherically symmetric with metric
\begin{eqnarray}
ds^2= - e^{2\Gamma}dt^2 + e^{2\Lambda}dr^2 + r^2(d\theta^2  + \sin^2\theta d\phi^2),
\end{eqnarray} 
where $\Gamma$ and $\Lambda$ depend on the radial coordinate $r$ only. The scalar fieds are static and the target space manifold is the round three-dimensional sphere
$\mathbb{S}^3$ with the metric 
\begin{eqnarray}
\gamma_{ab}d\varphi^a d\varphi^b= a^2\left[d\chi^2 + \sin^2\chi(d\varTheta^2 + \sin^2\varTheta d\Phi^2) \right], 
\end{eqnarray}
where $a>0$ is the radius of $\mathbb{S}^3$ and $\chi$, $\varTheta$ and $\Phi$ are the standard angular coordinate on $\mathbb{S}^3$. Since the scalar fields are time independent 
the map $\varphi : \text{\it space} \to \text{\it target space}$ reduces to a map $\varphi : \Sigma \to \mathbb{S}^3$ where $\Sigma$ is the spacial slice 
orthogonal to the timelike Killing vector field $\frac{\partial}{\partial t}$ and which is  diffeomorfic to $\mathbb{R}^3$. The explicit form of the map 
$\varphi : \Sigma \to \mathbb{S}^3$ is given by $(\chi=\chi(r), \Theta=\theta, \Phi=\phi)$. For this form of the $\varphi : \Sigma \to \mathbb{S}^3$ the dimensionally reduced equations are the following
\begin{eqnarray} 
   	&&\frac{2}{r}e^{-2\Lambda} \Lambda^{\prime} + \frac{1}{r^2}\left(1-e^{-2\Lambda}\right)=8\pi G_{*} A^{4}(\chi) {\tilde \varepsilon}
   	+a^2 \left(e^{-2\Lambda} {\chi^{\prime}}^2 + 2 \frac{\sin^2\chi}{r^2}\right)  + 2V(\chi), \label{DRE} \\ 
   	&&\frac{2}{r}e^{-2\Lambda} \Gamma^{\prime} - \frac{1}{r^2}\left(1-e^{-2\Lambda}\right)=8\pi G_{*} A^{4}(\chi) {\tilde p}
   	+a^2 \left(e^{-2\Lambda} {\chi^{\prime}}^2 - 2 \frac{\sin^2\chi}{r^2}\right)  - 2V(\chi), \\
   	&&\chi^{\prime\prime} + \left(\Gamma^\prime - \Lambda^\prime + \frac{2}{r}\right)\chi^{\prime}= \left[2\frac{\sin\chi\cos\chi}{r^2}  + \frac{1}{a^2} \frac{\partial V(\chi)}{\partial\chi} + \frac{4\pi G_{*}}{a^2} A^4(\chi)\frac{\partial \ln A(\chi)}{\partial\chi}(\tilde{\varepsilon} - 3{\tilde p})\right]e^{2\Lambda},\\
   	&& {\tilde p}^\prime = - (\tilde{\varepsilon} + {\tilde p}) \left[\Gamma^\prime + \frac{\partial \ln A(\chi)}{\partial\chi} \chi^\prime \right], \label{HSE}
\end{eqnarray} 
where the prime denotes differentiation with respect to $r$.
   
Asymptotic flatness requires $\Gamma(\infty)=\Lambda(\infty)=0$ and $\chi(\infty)=k\pi$ with $k$ being integer ($k \in \mathbb{Z}$). Without loss of generality we shall put $k=0$.  Regularity at the center requires $\Lambda(0)=0$ and $\chi(0)=n\pi$ where $n\in \mathbb{Z}$. 
The  integer  $n$ is in fact the  topological charge of the neutron star. Since $\chi(\infty)=0$ the map $\varphi :\Sigma\to \mathbb{S}^3$
can naturally be extended to  a map $\varphi :\Sigma\cup\infty \to \mathbb{S}^3$. Taking into account that    $\Sigma\cup \infty = \mathbb{R}^3\cup\infty$ is topologically $\mathbb{S}^3$ we have an effective map $\varphi : \mathbb{S}^3\to \mathbb{S}^3$. The topological charge $n$ is just the degree of the map $\varphi : \mathbb{S}^3\to \mathbb{S}^3$ given by 
\begin{eqnarray}
\deg\varphi=\int_{\Sigma}\varphi_*Vol_{\mathbb{S}^3},
\end{eqnarray}
where $\varphi_*Vol_{\mathbb{S}^3}$ is a 3-form on $\Sigma$ which is the  pull-back  of the normalized volume form  $Vol_{\mathbb{S}^3}$ on the target space $\mathbb{S}^3$ (i.e. $\int_{\mathbb{S}^3} Vol_{\mathbb{S}^3}=1$). A direct calculation indeed shows that $\deg\varphi=n$.
   
The system of reduced field equations  (\ref{DRE})--(\ref{HSE}) supplemented with the equation of state of the baryonic matter ${\tilde p}={\tilde p}({\tilde \varepsilon})$, with the above mentioned asymptotic and regularity conditions as  well as with a specified central energy density ${\tilde \varepsilon}_{c}$, describes the structure of the topological  neutron stars.

\section{Perturbations of the background solutions}

We consider time dependent radial linear perturbations over the spherically symmetric and static
background of topological neutron stars obtained after solving the reduced system of equations (\ref{DRE})--(\ref{HSE}). 
The perturbations of the spacetime metric is described by the functions $\delta \Lambda=\delta\Lambda(t,r)$ and $\delta \Gamma=\delta\Gamma(t,r)$ while the perturbation of the 
scalar field is described by $\delta\chi=\delta\chi(t,r)$, namely
\begin{eqnarray}
&&ds^2 = - e^{2\Gamma_0 + 2\delta\Gamma}dt^2 + e^{2\Lambda_0 + 2 \delta\Lambda}dr^2 + r^2(d\theta^2 + \sin^2\theta d\phi^2),\\
&&\chi = \chi_0 + \delta\chi.
\end{eqnarray}  
From now on the subscript ``0'' will refer to the quantities associated with the background solution. The radial perturbations of the fluid are described as usual by the Lagrangian
displacement $\zeta=\zeta(t,r)$. After tedious calculations one can show that the perturbations of the field equations result in two
coupled, second order wave  equations for $\zeta$ and $\delta\chi$, namely  
\begin{eqnarray}
&&({\tilde \varepsilon}_{0} + {\tilde p}_{0}) e^{2\Lambda_{0}-2\Gamma_{0}}\partial^2_{t}\zeta  + ({\tilde \varepsilon}_{0} + {\tilde p}_{0})\partial_{r}\delta\Gamma + 
\left[\partial_{r}\Gamma_{0} + \alpha(\chi_{0})\partial_{r}\chi_{0}\right](\delta {\tilde \varepsilon} + \delta {\tilde p})  \nonumber \\ 
&&+ \partial_{r}\delta {\tilde p} +
\alpha(\chi_0) ({\tilde \varepsilon}_{0} + {\tilde p}_{0}) \partial_{r}\delta\chi  + {\tilde \beta}(\chi_0) ({\tilde \varepsilon}_{0} + {\tilde p}_{0}) \partial_{r}\chi_0\delta\chi
=0 , \label{eq:PertEq1}\\
\nonumber\\
&&-e^{-2\Gamma_{0}} \partial^2_{t} \delta \chi + e^{-2\Lambda_{0}} \partial^2_{r} \delta \chi + e^{-2\Lambda_{0}}\left[\partial_{r}\Gamma_{0} - \partial_{r}\Lambda_{0} + \frac{2}{r} \right] \partial_{r}\delta\chi + e^{-2\Lambda_{0}}\partial_{r}\chi_{0}\left[\partial_{r}\delta\Gamma - \partial_{r}\delta\Lambda\right]  \nonumber \\
&&+ \left[-\frac{2}{r^2}\sin(2\chi_0) + \frac{2}{a^2} \partial_{\chi}V(\chi_0) 
- 8\pi G_{*} \frac{\alpha(\chi_0)}{a^2} A^4(\chi_{0})({\tilde \varepsilon}_{0} -3 {\tilde p}_{0})\right]\delta\Lambda \nonumber \\
&&-\left[\frac{2}{r^2}\cos(2\chi_0) + \frac{2}{a^2} \partial^2_{\chi}V(\chi_0) 
+ 4\pi G_{*} \frac{{\tilde \beta}(\chi_0)}{a^2} A^4(\chi_{0})({\tilde \varepsilon}_{0} -3 {\tilde p}_{0}) + 
 16\pi G_{*} \frac{\alpha^2(\chi_0)}{a^2} A^4(\chi_{0})({\tilde \varepsilon}_{0} -3 {\tilde p}_{0})\right]\delta\chi \nonumber\\
&& -  4\pi G_{*} \frac{\alpha(\chi_0)}{a^2} A^4(\chi_{0})(\delta {\tilde \varepsilon} -3 \delta{\tilde p}) =0 , \label{eq:PertEq2}
\end{eqnarray}
where $\alpha(\chi)=\frac{d\ln A(\chi)}{d\chi}$ and $\beta(\chi)=\frac{d^2\ln A(\chi)}{d\chi^2}$. The perturbations of the metric functions, the energy density and the pressure  are expressed in terms of $\zeta$ and $\delta\chi$, and their first derivative in 
the radial coordinate as follows:   
\begin{eqnarray}
&&\delta \Lambda = a^2 r\partial_{r}\chi_0 \delta\chi - 4\pi G_{*} A^4(\chi_0) ({\tilde \varepsilon}_{0} + {\tilde p}_{0}) e^{2\Lambda_0} r\zeta ,\\
\nonumber \\
&&\delta {\tilde \varepsilon}= - ({\tilde \varepsilon}_{0} + {\tilde p}_{0})\left[\frac{e^{-\Lambda_0}}{r^2} \partial_{r}\left(e^{\Lambda_0} r^2\zeta \right) + \delta\Lambda\right] \\
&& \hspace{1.3cm} -\left[ \partial_{r}{\tilde \varepsilon}_{0} + 3\alpha(\chi_0) ({\tilde \varepsilon}_{0} + {\tilde p}_{0})\partial_{r}\chi_0\right]\zeta
- 3\alpha(\chi_0)({\tilde \varepsilon}_{0} + {\tilde p}_{0}) \delta\chi , \\ 
\nonumber \\
&&\delta {\tilde p}= {\tilde c}^2_{s} \delta {\tilde \varepsilon}, \\ 
\nonumber \\
&&\partial_{r}\delta\Gamma= \frac{1}{r}\left[1 + r^2\left(8\pi G_* A^4(\chi_0){\tilde p}_{0} - 2V(\chi_0) \right)\right]e^{2\Lambda_0} \delta\Lambda \\
&& \hspace{1.3cm} + re^{2\Lambda_0}\left[-\partial_{\chi}V(\chi_0) + 16\pi G_* \alpha(\chi_0) A^{4}(\chi_0){\tilde p}_0 \right]\delta \chi \nonumber \\
&& \hspace{1.3cm} +  a^2 r \partial_{r}\chi_0 \partial_{r}\delta\chi + 4\pi G_{*} e^{2\Lambda_0} r A^{4}(\chi_0) \delta {\tilde p} .
\end{eqnarray}
Here ${\tilde c}^2_{s} $ is the Jordan frame sound velocity defined by ${\tilde c}^2_{s} =\frac{d{\tilde p}}{d{\tilde \varepsilon}}$.

Let us discuss the boundary conditions that have to be impose on $\zeta$ and $\delta\chi$. The regularity of the energy density and
pressure perturbations at the center of the star requires $\zeta(t,r=0)=0$. The perturbation of the scalar field $\chi$ has to be zero at the star center, 
$\delta\chi(t,r=0)=0$. The boundary condition at the star surface is that the Lagrangian perturbation of the  pressure $\Delta {\tilde p}$ vanishes. In explicit form we have 
\begin{eqnarray}\label{eq:BC_surface_deltaP}
&&\Delta {\tilde p}(t,r=R)= - {\tilde c}^2_s ({\tilde \varepsilon}_{0} + {\tilde p}_{0}) \left[\frac{e^{\Gamma_0}}{r^2}\partial_{r}\left(e^{-\Gamma_0}r^2\zeta\right) +  \right.  \left(a^2 r\partial_{r}\chi_0  + 3\alpha(\chi_0)\right)\partial_{r}\chi_0 \zeta  \\
&&\hskip 5cm \left.
 +\left(a^2 r\partial_{r}\chi_0  + 3\alpha(\chi_0)\right) \delta\chi\Big]\right\vert_{r=R}=0 . \nonumber
\end{eqnarray} 

Only the perturbation $\delta\chi$ of the scalar field can propagate outside the star and $\delta\chi$ has to satisfy the radiative (outgoing)
asymptotic conditions, namely
\begin{eqnarray}
\partial_{t} (r\delta\chi) + \partial_{r}(r\delta\chi)=0 .
\end{eqnarray}
 
 \section{Results}
 \subsection{Background solutions}
  \begin{figure}
 	\includegraphics[width=0.45\textwidth]{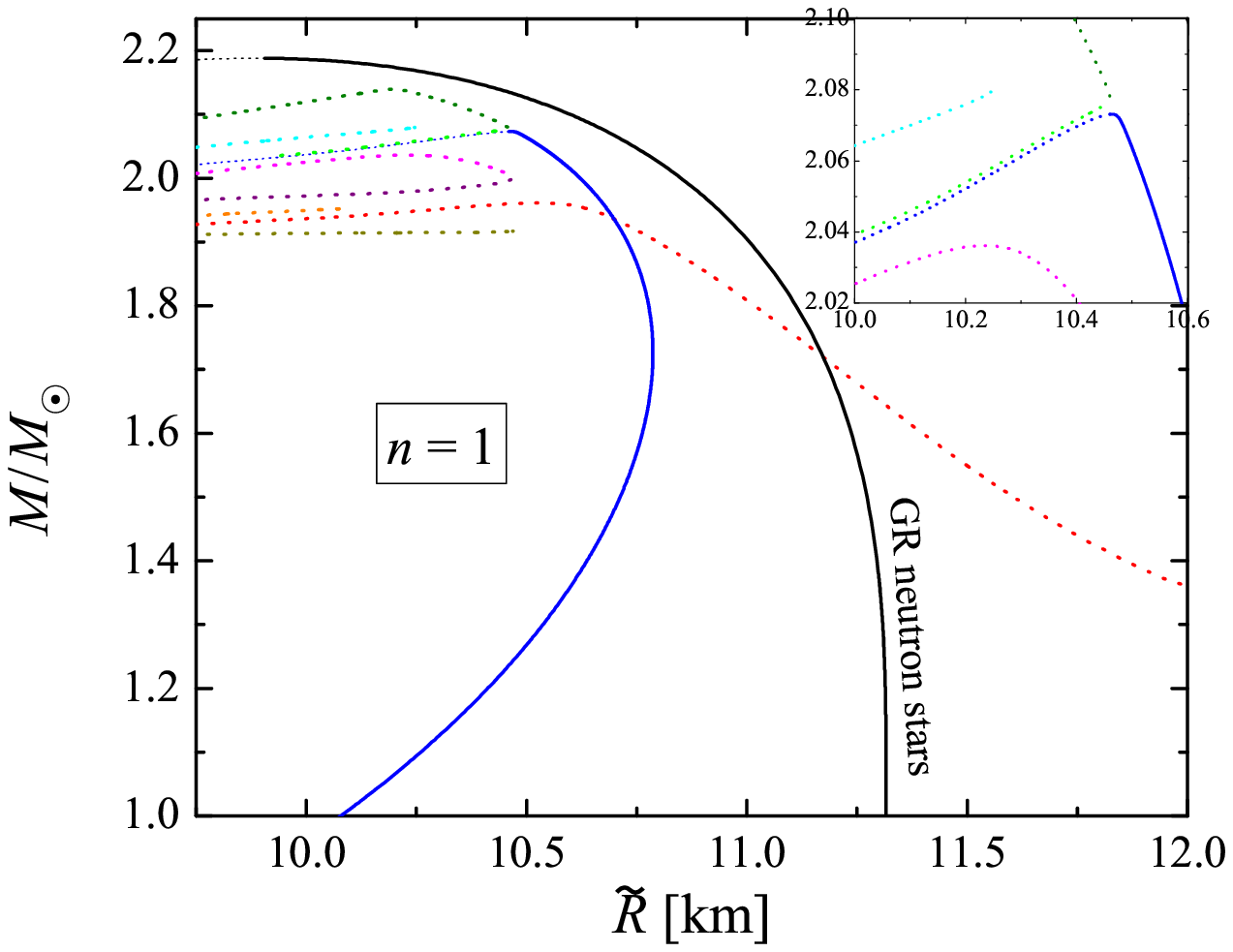}
 	\includegraphics[width=0.45\textwidth]{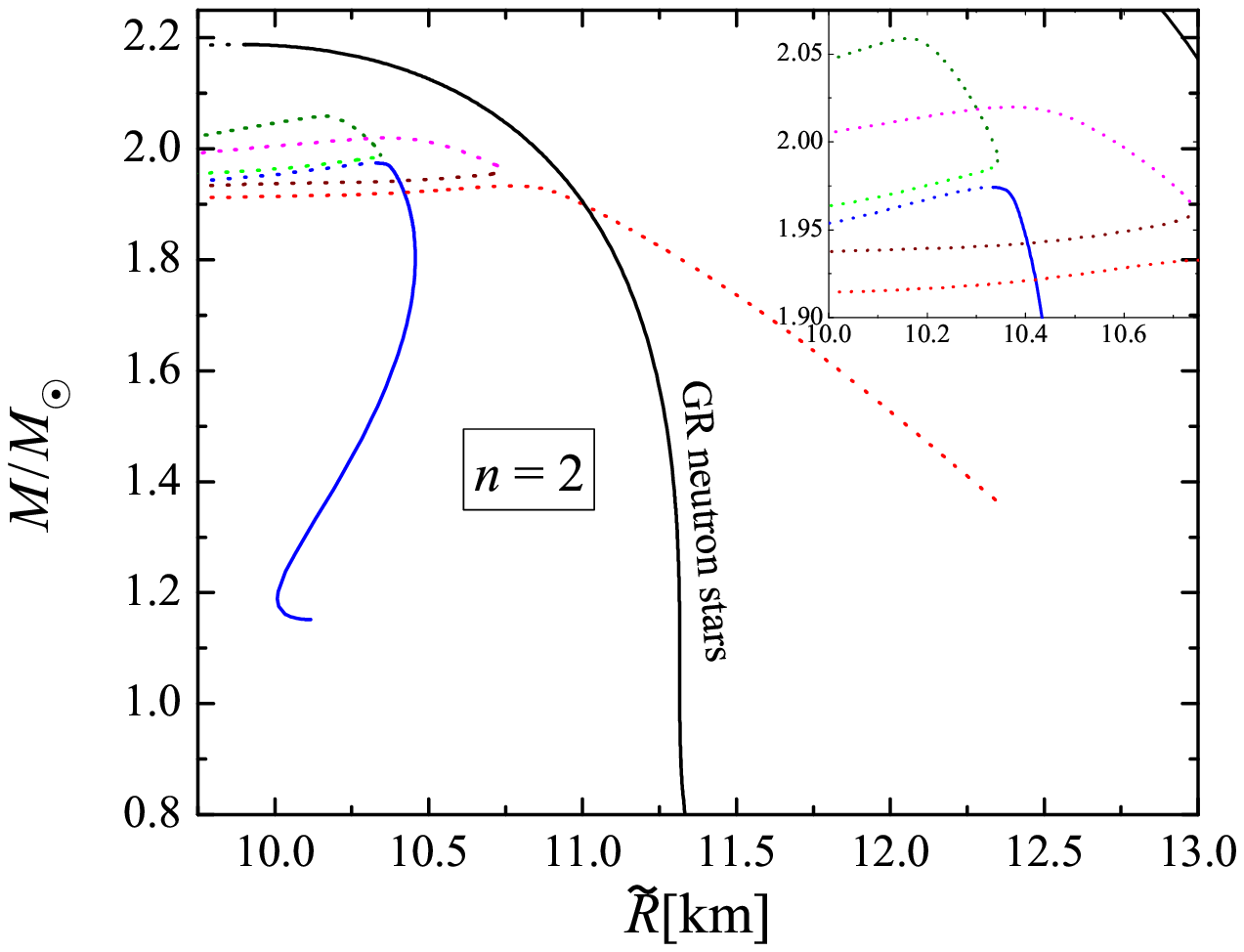}
 	\caption{The mass as a function of radius for neutron star models with $n=1$ (left panel) and $n=2$ (right panel). The free parameters are $\beta=0.08$ and $a^2=0.001$. The GR sequence of models is depicted with black line, while the TMST branches -- with lines with different colors. The stable part of the branches is plotted with a solid line while the unstable part of the branches -- with dotted line. A zoom of the region where several new additional branches appear is given in the top-right corner of each graph.}
 	\label{fig:MR_PI_2PI}
 \end{figure}

First let's briefly discuss the equilibrium neutron star solutions that we will perturb. We will focus on the models presented in \cite{Doneva:2019ltb} in order to be able to perform a better comparison. The conformal factor is chosen to be
\begin{equation}\label{eq:conformal_factor}
A(\varphi)=\exp(\beta \sin^2 \chi),
\end{equation}
where $\beta$ is a parameter. The EOS we employ is APR4 \cite{AkmalPR,Read2009}. The presented results will be for $\beta=0.08$ and $a^2=10^{-3}$. As we will discuss below, though, the stability is tested for a  larger range of parameters and conformal factors. 
 
In Fig. \ref{fig:MR_PI_2PI} the mass as a function of radius is presented for $n=1$ and $n=2$. In both cases we have two branches spanning to small masses (the blue and the red lines) and many additional branches appearing at larger masses. The GR sequence of neutron stars is also plotted for a comparison. The style of the lines is connected with their stability and will be commented in the next subsection. Even though it might seem that the different branches at larger $M$ are connected, if one looks in the zoom in the top-right corner, it is evident that there is a clear distinction between the lines with different colors. For more details we refer the reader to \cite{Doneva:2019ltb} where the neutron star sequences are discussed thoroughly.

The differences between the GR and TMST models tends to be larger for small masses, while for large masses at least some of the TMST branches get closer to the GR one. It is very interesting and observationally important that the scalar charge for these models, defined as the coefficient in front of the $1/r$ scalar field asymptotic at infinity, is zero \cite{Doneva:2019ltb}. Therefore, there is no dipole scalar gravitational radiation and the binary pulsar observations can not impose constraints on the theory.
 
Even though a large number of branches exists, most of them show undesired peculiarities as discussed in \cite{Doneva:2019ltb}. For example some of the solutions belonging to the red line branch and the branches appearing at larger masses have peculiar behavior as commented in \cite{Doneva:2019ltb} that  was interpreted as a sign of instability. Only the blue line branch behaves well (at least until the maximum of the mass is reached) and in addition, the neutron star radius is within the observational constraints.

Even though the $n=1$ and $n=2$ cases lead to different scalar field boundary conditions, the overall picture of the branches looks quite similar. In both cases we have two branches that span to smaller masses and a number of high mass branches. The number of these additional hight mass branches is different for $n=1$ and $n=2$, but as we discussed, they are supposedly unstable and we will not comment on them further. The $n=2$ case leads as well to somewhat larger differences from GR in most cases. This is expected since the $n=2$  neutron stars have in general stronger scalar field -- at the center $\chi(0)=2\pi$ compared to $\chi(0)=\pi$ in the $n=1$ case. 

The supposedly stable blue line $n=2$ sequence is terminated at some finite mass meaning that no stable $n=2$ solutions can exist below it. In addition, the maximum of the mass for $n=2$ is slightly below two solar masses. This is an expected behavior for topological solutions -- they exist only in a fixed domain of the parameters space. Still for very large masses $n=1$ neutron stars can exist as well the zero scalar field ($n=0$) GR  solutions. Therefore, the considered TMST is not in a contradiction with the observations.

 \subsection{Radial perturbations} 
It turns out that it is numerically difficult to impose the boundary condition at the stellar surface in the form of eq. \eqref{eq:BC_surface_deltaP}. The output signal is highly dependent on the proper implementation of this condition and the precision that is used. This is an undesired behavior because it can lead to numerical instabilities in certain cases and needs fine tuning. That is why we used a standard approach adopted from pure GR \cite{RuoffPhD}, that is to introduce a new dimensionless function 
 \begin{equation}
 Z=({\tilde \varepsilon_0} + {\tilde p_0})r\zeta e^{2\Lambda_0}.
 \end{equation}
We are dealing with neutron stars where both  ${\tilde \varepsilon_0}$ and  ${\tilde p_0}$ vanish at the stellar surface, and therefore the perturbation $Z$ will vanish as well. Clearly a zero boundary condition at the surface is much more robust and that is why we will work with the $Z$ function.
 
 We have performed a time evolution of the perturbation equations \eqref{eq:PertEq1} and \eqref{eq:PertEq2}, where the perturbation variable $\zeta$ is transformed to $Z$. Since this is a straightforward calculation we will not give here the perturbation equations explicitly. A Gauss pulse for $\delta\chi$, located at several neutron star radii outside the star and having zero velocity at $t=0$, is used as initial data and is evolved in time. At $t=0$ we have imposed zero perturbation $Z$. Since $Z$ can be nontrivial only inside the star, it will remain zero until the  scalar field perturbation reaches the star. From that moment on  $\delta\chi$ will excited $Z$ because the two perturbation equations are coupled. We perform the time evolution for roughly 50ms and make a Fourier transformation of the $\delta\chi$ signal at roughly ten neutron star radii. The validity of our code is counterchecked against different particular cases and shows very good agreement -- the pure GR case \cite{Kokkotas:2000up,RuoffPhD} and the results for scalarized neutron stars in STT \cite{Mendes:2018qwo}. In addition, the radial fluid modes frequencies extracted from $\delta\chi$ and $Z$ coincide.
 
Let us comment on the main differences between the signal for GR models and models with nonzero scalar field. As far as radial perturbations are considered and neglecting viscous effects in the neutron star interior, the $Z$ perturbation inside the star will not be damped in the pure GR case since no gravitational radiation can be emitted. This is not the case in the presence of scalar field, since its perturbation carries away energy at infinity which leads to damping of the neutron star radial fluid modes. This was observed already in \cite{Mendes:2018qwo} and applies to our problem as well. 

After examining the radial perturbations of several solutions from each branch in Fig. \ref{fig:MR_PI_2PI} we could conclude that the radial stability/instability is indeed as predicted in \cite{Doneva:2019ltb}. The stable part of the branches is depicted with solid lines in Fig. \ref{fig:MR_PI_2PI}, while dotted line denotes instability. For both cases $n=1$ and $n=2$, the only stable branch is the one depicted by blue line and only up to its maximum of the mass, while the rest of the branches are unstable. As a matter of fact this is the only ``nicely behaving'' branch reaching as well to low NS masses and having radii of the same order as the GR case for the whole range of astrophysically relevant masses. Moreover, as pointed out in \cite{Doneva:2019ltb}, the radial profiles of the metric functions, the scalar field and the pressure for models belonging to the blue line branch (before a maximum of the mass is reached), are nicely behaving. As expected, the results show that after reaching a maximum of the mass the neutron stars become unstable.

 \begin{figure}
	\includegraphics[width=0.45\textwidth]{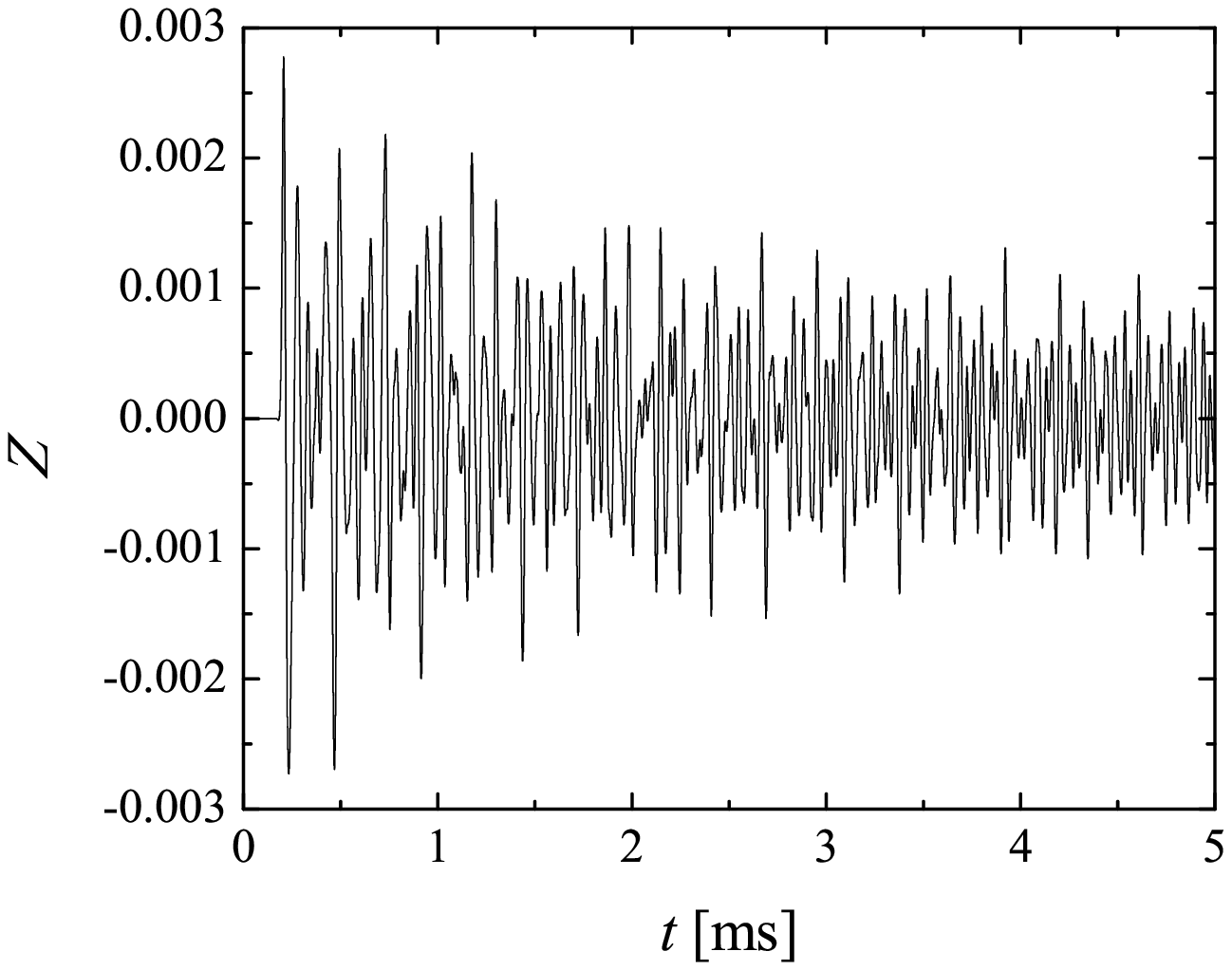}
	\includegraphics[width=0.45\textwidth]{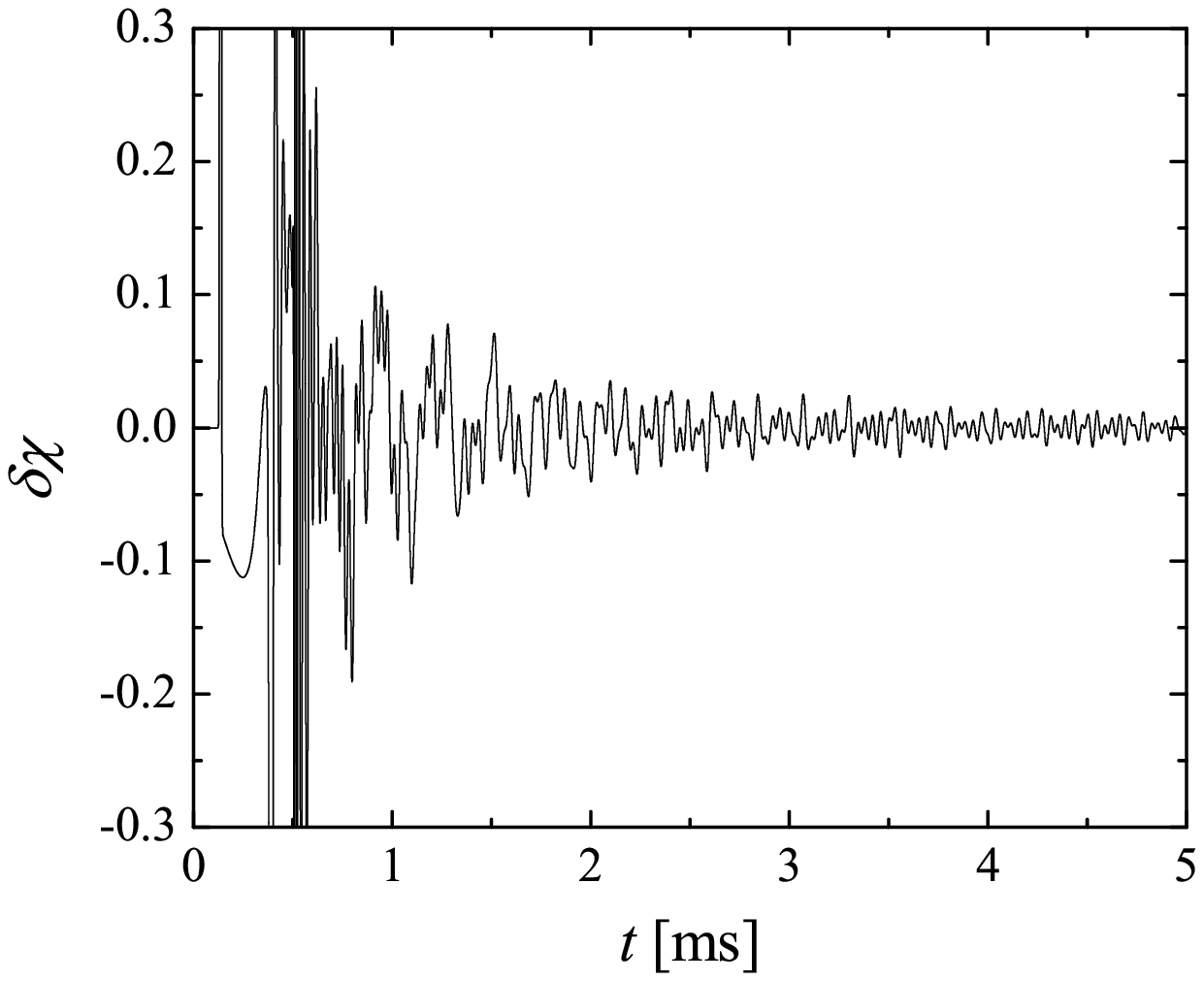}
	\caption{The time evolution of $Z$ and $\delta\chi$ for TMST neutron stars belonging to the blue line branch in Fig. \ref{fig:MR_PI_2PI} with $n=1$, $\beta=0.08$ and $a^2=0.001$. The density is chosen to be before the maximum of the mass is reached $\varepsilon_{c}=1.28 \times 10^{15} {\rm g/cm^3}$ with mass $M=1.691 M_\odot$ which leads to a stable solution. The $Z$ perturbation is extracted in the neutron star interior while the $\delta\chi$ signal is observed at roughly ten neutron star radii away from the star.}
	\label{fig:zeta_dchi_2_4}
\end{figure}

\begin{figure}
	\includegraphics[width=0.45\textwidth]{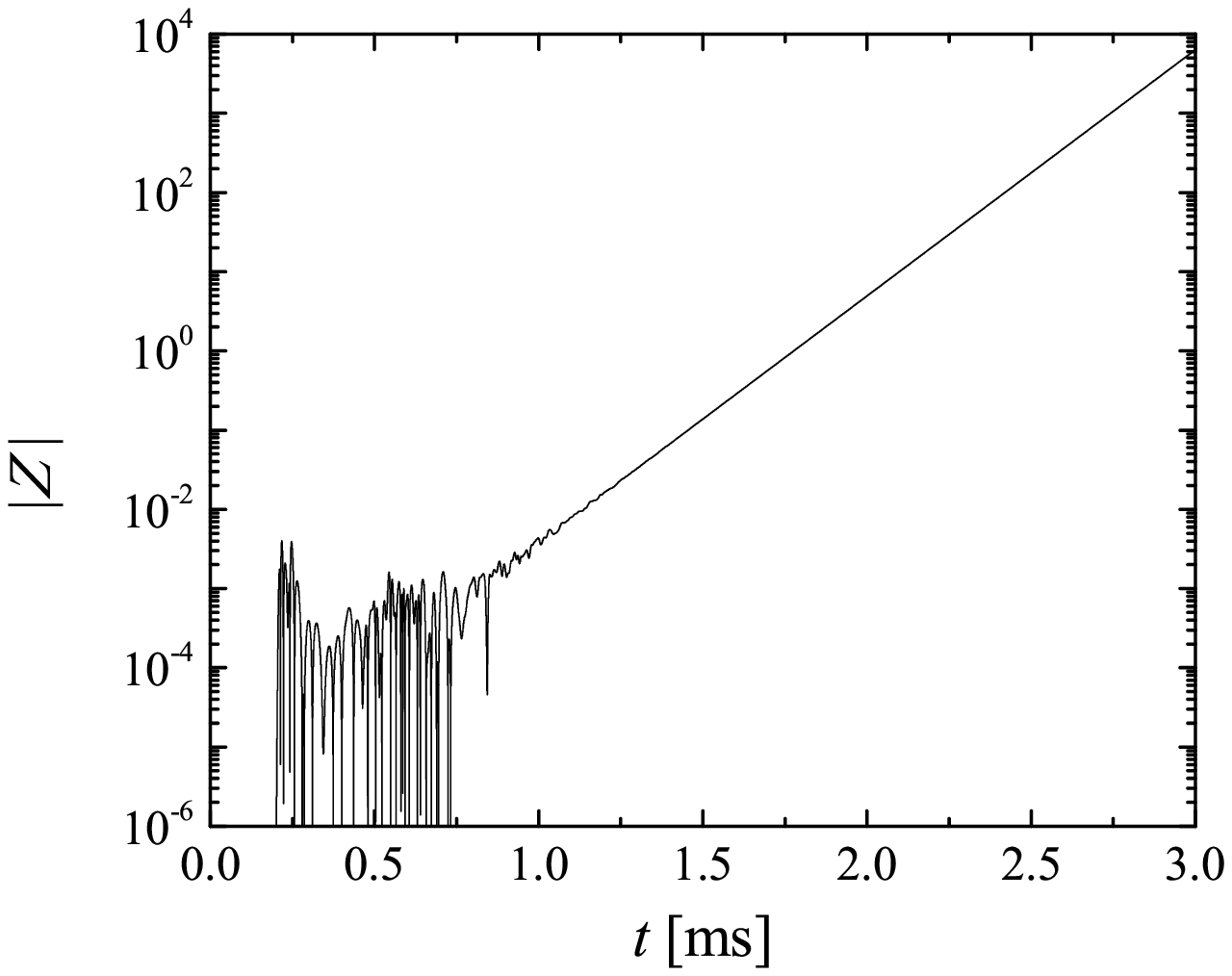}
	\includegraphics[width=0.45\textwidth]{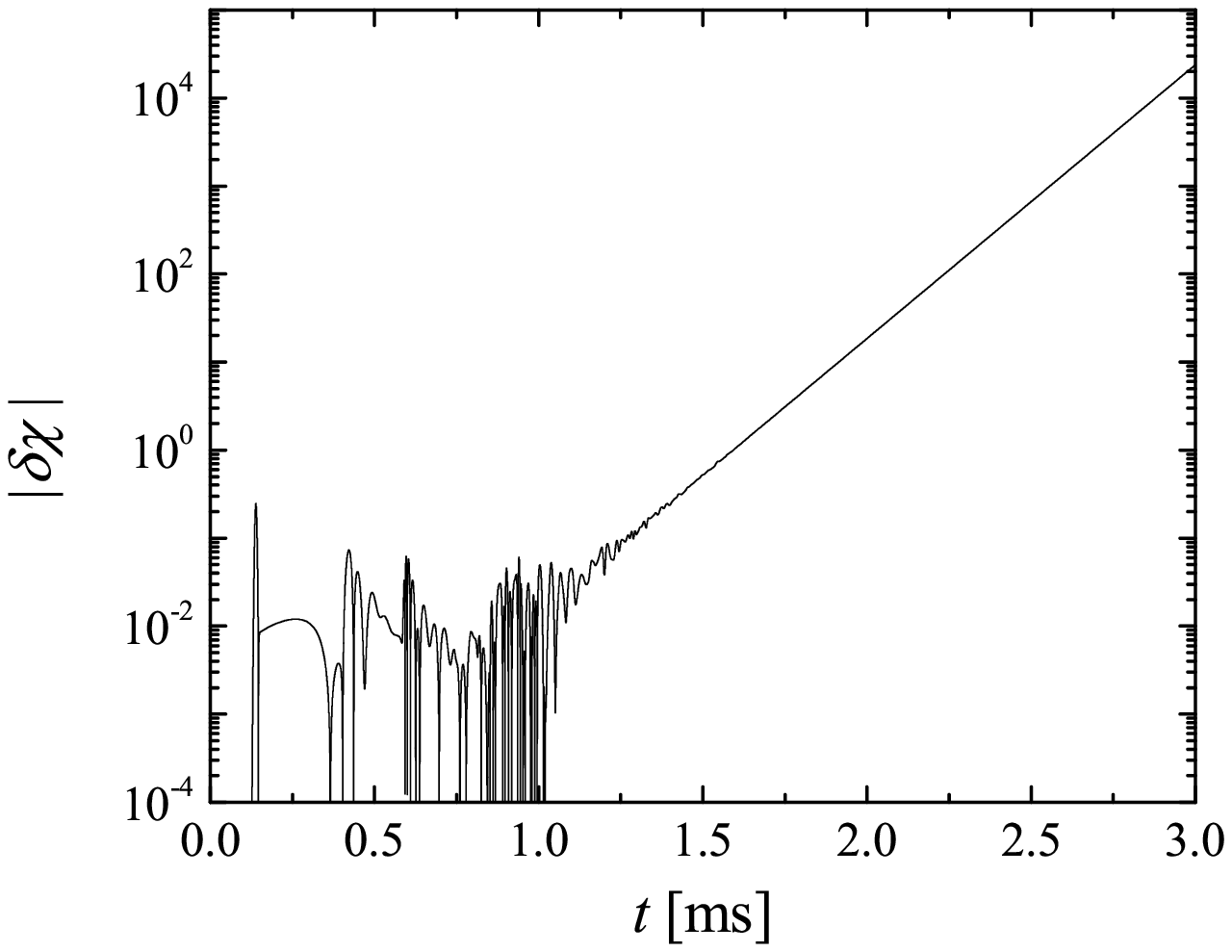}
	\caption{The time evolution of $Z$ and $\delta\chi$ for TMST neutron stars belonging to the blue line branch in Fig. \ref{fig:MR_PI_2PI} with $n=1$, $\beta=0.08$ and $a^2=0.001$. The density is chosen to be after the maximum of the mass is reached $\varepsilon_{c}=2.07 \times 10^{15} {\rm g/cm^3}$ with mass $M=2.059 M_\odot$ which leads to a unstable solution. The $Z$ perturbation is extracted in the neutron star interior while the $\delta\chi$ signal is observed at roughly ten neutron star radii away from the star. The $y$ axis is in a logarithmic scale. }
	\label{fig:zeta_dchi_2_7}
\end{figure}

Exemplary waveforms for models belonging to the blue line branch in Fig. \ref{fig:MR_PI_2PI} (for $n=1$) are presented in Fig. \ref{fig:zeta_dchi_2_4} and Fig. \ref{fig:zeta_dchi_2_7}. In the former figure a TMST neutron star before the maximum of the mass is chosen and therefore it is stable while in latter figure -- a models after the maximum of the mass and thus unstable. The behavior of the unstable neutron star radial modes belonging to the rest of the branches is similar -- after several oscillations the perturbation starts to increase exponentially.

Let us examine in more detail the stable model signal in Fig. \ref{fig:zeta_dchi_2_4} . As one can see, $Z$ is zero before the scalar field perturbation ``hits'' the star. Afterwards $Z$ oscillates and slowly decreases in time contrary to what is observed in the GR case. The reason is the coupling with the scalar field and the loss of energy through scalar gravitational radiation. As expected, the amplitude of $Z$ is much smaller than $\delta\chi$ due to the fact that it is only excited through the coupling between the two perturbations. 

The appearance of scalar modes in addition to the modes related to the radial oscillations of the fluid is observed in all of the studied cases. This is natural to expect for the system of perturbed equations we consider \cite{1986GReGr..18..913K} and given as well the analogy with the spacetime $w$-modes \cite{1992MNRAS.255..119K}. Scalar modes were found as well  in the standard STT \cite{Mendes:2018qwo}. The best visualization of the phenomenon is in the scalar field perturbation depicted in right panel of Fig. \ref{fig:zeta_dchi_2_7} because of the logarithmic scale (even though this is an unstable model and the instability develops after several oscillation). In the very first part of the signal several highly damped oscillations are observed that quickly disappear followed in some cases by an exponential tail and afterwards the signal is dominated by the modes related to the radial oscillations of the fluid. We will not explore this problem in more details due to the following reason. The main goal of our paper is to check the stability of different TMST neutron star branches while extracting the scalar modes is extremely difficult task because of the small damping time. Based on our tests, we can conclude that in most cases the scalar modes of topological TMST neutron stars are much more difficult to be extracted compared to pure STT and there is a huge error in the determined frequencies due to the small number of scalar mode oscillations observed in the signal. There might be different scenarios leading to stronger and better pronounced scalar modes, such as the gravitational collapse, and we plan to address this problem in the future.

\begin{figure}
	\includegraphics[width=0.45\textwidth]{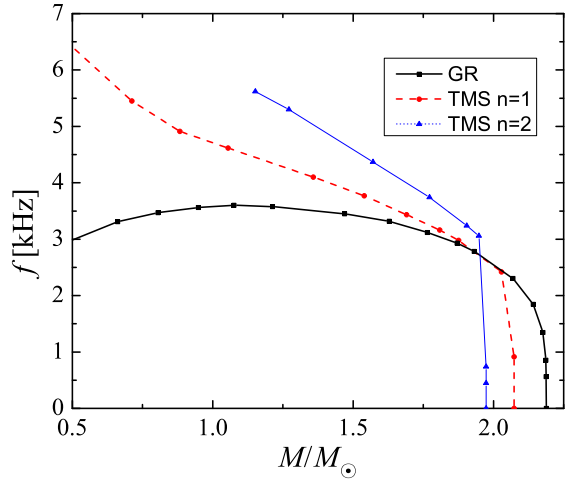}
	\caption{The fundamental radial $F$ mode frequency as a function of the neutron star mass for GR models and TMST neutron stars belonging to the stable blue line branch in Fig. \ref{fig:MR_PI_2PI}. The cases of both $n=1$ and $n=2$ are shown, while $\beta=0.08$ and $a^2=0.001$.}
	\label{fig:Freq_M}
\end{figure}

After performing a Fourier transformation of the observed signal, several peaks appear in the spectrum that can be identified with the fist radial mode F (which have no nodes inside the star) and its overtones. A clear peak corresponding to the scalar modes can be rarely observed. The frequencies of the F mode are plotted in Fig. \ref{fig:Freq_M} for pure GR and TMST neutron stars belonging to the blue line branch in Fig. \ref{fig:MR_PI_2PI} both for $n=1$ and $n=2$. As one can see, the largest difference in comparison with the Einstein's theory is for small masses similar to the behavior of the equilibrium models. As the mass increases the frequencies start to rapidly decrease until a zero frequency is reached. This happens at the maximum of the mass for the corresponding branch and signals the onset of instability. 

The qualitative behavior of the $n=1$ and $n=2$ cases is quite similar. The main difference is as expected from the study of the background models -- the $n=2$ modes reach zero frequency at smaller masses compared to $n=1$  and the $n=2$  sequence is terminated at some finite small value of the mass where the solutions disappear. The F mode frequencies of the intermediate and smaller mass models are larger than GR and increase with the increase of $n$.

The results presented so far are for a particular conformal factor \eqref{eq:conformal_factor} and values of $\beta$ and $a^2$. We have performed more simulations in order to check up to what extend the conclusions we have made are universal. The difficult part is that topological solutions exist only for certain ranges of parameters and moreover -- finding the whole spectrum of solutions for a given conformal factor and fixed values of $\beta$ and $a^2$ is an extremely time-consuming task. We have considered the conformal factor \eqref{eq:conformal_factor} and several different values of $\beta$, as well as the conformal factor $A(\varphi)=\exp(\beta \chi^2)$. In all cases we have found that no more than one stable branch can exist for each value of the topological charge $n$ and all the additional branches are unstable. Therefore, the conclusions we have made in this chapter seem to be more or less universal for all of the topological neutron stars in the considered class of TMST.

\section{Conclusions}
In the present paper we have examined the stability against radial perturbations of topological neutron stars in a class of tensor-multi-scalar theories of gravity that are new and very interesting solutions due to the presence of nonzero topological charge. These solutions have zero scalar charge and thus no scalar dipole radiation can be emitted. Contrary to the pure GR case, for a fixed value of the topological charge many branches of solutions can exist. That is why it was very important to identify which are the stable branches that can realize in practice and lead to astrophysically relevant neutron stars. 

The equations governing the radial perturbations were derived and they reduce to a system of two coupled wave-type equations -- one for the Lagrangian displacement of the fluid and one for the perturbation of the scalar field. The radial modes are calculated  via time evolution of the system of equations. The results show that for a fixed topological charge only one of the neutron star branches is stable while all the rest are unstable. Moreover, this is exactly the stable branch that spans a wide range of neutron star masses and have radii that are in the preferred range according to the observations. Similar to the GR case, this branch looses stability after the maximum of the mass. It is interesting to note, that these conclusions coincide with the rough analysis of the solutions performed in \cite{Doneva:2019ltb} where it was demonstrated that most of the topological neutron star branches show undesired peculiarities which is a signal of instability. 

We have examined as well the frequencies of the radial fluid modes for different values of the topological charge. The differences with GR are small for the more massive models and increase for smaller mass neutron stars.  The radial mode frequencies are generally larger for larger value of the topological charge. For higher masses close to the maximum one the frequencies star to rapidly decrease reaching zero at the maximum of the mass that signals the onset of instability.

Let us comment on the possible astrophysical applications and future prospects. Having only one stable sequence of solutions for every value of the topological charge is a nice and desired property. Thus, it should be  possible to come up with a strategy to observationally distinguish between neutron stars with different topological charge and thus, put constraints on the theory. This would not be possible solely on the basis of the mass and radius observations, because even though the differences with the GR solutions can be large, the uncertainties in the nuclear matter EOS are comparable. Another trilling and more promising possibility is to use the gravitational wave observations. Contrary to GR where the radial perturbations can not lead to gravitational wave emission, in the considered TMST, breathing modes and gravitational wave echoes can be observed. Similar to pure STT, though, this is possible only for a certain form of the conformal factors for which $\alpha(\chi)=\frac{d\ln A(\chi)}{d\chi}$
is linear for small $\chi$ \cite{Damour1992,Gerosa:2016fri}. In additional, the time evolution showed the appearance of scalar modes with very short damping time that can have possible imprints in the electromagnetic and gravitational wave emission \cite{1986GReGr..18..913K,1992MNRAS.255..119K,Mendes:2018qwo}. A detailed study of these phenomena including the observational prospect is a study underway and will be reported elsewhere.

\section*{Acknowledgements}
DD acknowledges financial support via an Emmy Noether Research Group funded by the German Research Foundation (DFG) under grant
no. DO 1771/1-1. DD is indebted to the Baden-Wurttemberg Stiftung for the financial support of this research project by the Eliteprogramme for Postdocs.  SY would like to thank the University of Tuebingen for the financial support.  This work was supported by DFG grant 413873357.
The partial support by the Bulgarian NSF Grant DCOST 01/6 and the  Networking support by the COST Actions  CA16104 and CA16214 are also gratefully acknowledged.


\bibliography{references}

\begin{thebibliography}{32}
\expandafter\ifx\csname natexlab\endcsname\relax\def\natexlab#1{#1}\fi
\expandafter\ifx\csname bibnamefont\endcsname\relax
  \def\bibnamefont#1{#1}\fi
\expandafter\ifx\csname bibfnamefont\endcsname\relax
  \def\bibfnamefont#1{#1}\fi
\expandafter\ifx\csname citenamefont\endcsname\relax
  \def\citenamefont#1{#1}\fi
\expandafter\ifx\csname url\endcsname\relax
  \def\url#1{\texttt{#1}}\fi
\expandafter\ifx\csname urlprefix\endcsname\relax\def\urlprefix{URL }\fi
\providecommand{\bibinfo}[2]{#2}
\providecommand{\eprint}[2][]{\url{#2}}

\bibitem[{\citenamefont{{Berti} et~al.}(2015)\citenamefont{{Berti}, {Barausse},
  {Cardoso}, {Gualtieri}, {Pani}, {Sperhake}, {Stein}, {Wex}, {Yagi}, {Baker}
  et~al.}}]{Berti2015a}
\bibinfo{author}{\bibfnamefont{E.}~\bibnamefont{{Berti}}},
  \bibinfo{author}{\bibfnamefont{E.}~\bibnamefont{{Barausse}}},
  \bibinfo{author}{\bibfnamefont{V.}~\bibnamefont{{Cardoso}}},
  \bibinfo{author}{\bibfnamefont{L.}~\bibnamefont{{Gualtieri}}},
  \bibinfo{author}{\bibfnamefont{P.}~\bibnamefont{{Pani}}},
  \bibinfo{author}{\bibfnamefont{U.}~\bibnamefont{{Sperhake}}},
  \bibinfo{author}{\bibfnamefont{L.~C.} \bibnamefont{{Stein}}},
  \bibinfo{author}{\bibfnamefont{N.}~\bibnamefont{{Wex}}},
  \bibinfo{author}{\bibfnamefont{K.}~\bibnamefont{{Yagi}}},
  \bibinfo{author}{\bibfnamefont{T.}~\bibnamefont{{Baker}}},
  \bibnamefont{et~al.}, \bibinfo{journal}{Classical and Quantum Gravity}
  \textbf{\bibinfo{volume}{32}}, \bibinfo{eid}{243001} (\bibinfo{year}{2015}),
  \eprint{1501.07274}.

\bibitem[{\citenamefont{{Doneva} and {Pappas}}(2018)}]{2018ASSL..457..737D}
\bibinfo{author}{\bibfnamefont{D.~D.} \bibnamefont{{Doneva}}} \bibnamefont{and}
  \bibinfo{author}{\bibfnamefont{G.}~\bibnamefont{{Pappas}}},
  \emph{\bibinfo{title}{{Universal Relations and Alternative Gravity
  Theories}}} (\bibinfo{year}{2018}), vol. \bibinfo{volume}{457} of
  \emph{\bibinfo{series}{Astrophysics and Space Science Library}}, p.
  \bibinfo{pages}{737}.

\bibitem[{\citenamefont{{Damour} and {Esposito-Farese}}(1993)}]{Damour1993}
\bibinfo{author}{\bibfnamefont{T.}~\bibnamefont{{Damour}}} \bibnamefont{and}
  \bibinfo{author}{\bibfnamefont{G.}~\bibnamefont{{Esposito-Farese}}},
  \bibinfo{journal}{Physical Review Letters} \textbf{\bibinfo{volume}{70}},
  \bibinfo{pages}{2220} (\bibinfo{year}{1993}).

\bibitem[{\citenamefont{Will}(2014)}]{Will:2014kxa}
\bibinfo{author}{\bibfnamefont{C.~M.} \bibnamefont{Will}},
  \bibinfo{journal}{Living Rev.\ Rel.} \textbf{\bibinfo{volume}{17}},
  \bibinfo{pages}{4} (\bibinfo{year}{2014}), \eprint{1403.7377}.

\bibitem[{\citenamefont{{Demorest} et~al.}(2010)\citenamefont{{Demorest},
  {Pennucci}, {Ransom}, {Roberts}, and {Hessels}}}]{Demorest10}
\bibinfo{author}{\bibfnamefont{P.~B.} \bibnamefont{{Demorest}}},
  \bibinfo{author}{\bibfnamefont{T.}~\bibnamefont{{Pennucci}}},
  \bibinfo{author}{\bibfnamefont{S.~M.} \bibnamefont{{Ransom}}},
  \bibinfo{author}{\bibfnamefont{M.~S.~E.} \bibnamefont{{Roberts}}},
  \bibnamefont{and} \bibinfo{author}{\bibfnamefont{J.~W.~T.}
  \bibnamefont{{Hessels}}}, \bibinfo{journal}{\nat}
  \textbf{\bibinfo{volume}{467}}, \bibinfo{pages}{1081} (\bibinfo{year}{2010}).

\bibitem[{\citenamefont{Antoniadis et~al.}(2013)\citenamefont{Antoniadis,
  Freire, Wex, Tauris, Lynch, van Kerkwijk, Kramer, Bassa, Dhillon, Driebe
  et~al.}}]{Antoniadis2013a}
\bibinfo{author}{\bibfnamefont{J.}~\bibnamefont{Antoniadis}},
  \bibinfo{author}{\bibfnamefont{P.~C.} \bibnamefont{Freire}},
  \bibinfo{author}{\bibfnamefont{N.}~\bibnamefont{Wex}},
  \bibinfo{author}{\bibfnamefont{T.~M.} \bibnamefont{Tauris}},
  \bibinfo{author}{\bibfnamefont{R.~S.} \bibnamefont{Lynch}},
  \bibinfo{author}{\bibfnamefont{M.~H.} \bibnamefont{van Kerkwijk}},
  \bibinfo{author}{\bibfnamefont{M.}~\bibnamefont{Kramer}},
  \bibinfo{author}{\bibfnamefont{C.}~\bibnamefont{Bassa}},
  \bibinfo{author}{\bibfnamefont{V.~S.} \bibnamefont{Dhillon}},
  \bibinfo{author}{\bibfnamefont{T.}~\bibnamefont{Driebe}},
  \bibnamefont{et~al.}, \bibinfo{journal}{Science}
  \textbf{\bibinfo{volume}{340}}, \bibinfo{pages}{1233232}
  (\bibinfo{year}{2013}).

\bibitem[{\citenamefont{{Damour} and {Esposito-Farese}}(1992)}]{Damour1992}
\bibinfo{author}{\bibfnamefont{T.}~\bibnamefont{{Damour}}} \bibnamefont{and}
  \bibinfo{author}{\bibfnamefont{G.}~\bibnamefont{{Esposito-Farese}}},
  \bibinfo{journal}{Classical and Quantum Gravity}
  \textbf{\bibinfo{volume}{9}}, \bibinfo{pages}{2093} (\bibinfo{year}{1992}).

\bibitem[{\citenamefont{Horbatsch et~al.}(2015)\citenamefont{Horbatsch, Silva,
  Gerosa, Pani, Berti, Gualtieri, and Sperhake}}]{Horbatsch:2015bua}
\bibinfo{author}{\bibfnamefont{M.}~\bibnamefont{Horbatsch}},
  \bibinfo{author}{\bibfnamefont{H.~O.} \bibnamefont{Silva}},
  \bibinfo{author}{\bibfnamefont{D.}~\bibnamefont{Gerosa}},
  \bibinfo{author}{\bibfnamefont{P.}~\bibnamefont{Pani}},
  \bibinfo{author}{\bibfnamefont{E.}~\bibnamefont{Berti}},
  \bibinfo{author}{\bibfnamefont{L.}~\bibnamefont{Gualtieri}},
  \bibnamefont{and} \bibinfo{author}{\bibfnamefont{U.}~\bibnamefont{Sperhake}},
  \bibinfo{journal}{Class.\ Quant.\ Grav.} \textbf{\bibinfo{volume}{32}},
  \bibinfo{pages}{204001} (\bibinfo{year}{2015}), \eprint{1505.07462}.

\bibitem[{\citenamefont{Yazadjiev and Doneva}(2019)}]{Yazadjiev:2019oul}
\bibinfo{author}{\bibfnamefont{S.~S.} \bibnamefont{Yazadjiev}}
  \bibnamefont{and} \bibinfo{author}{\bibfnamefont{D.~D.}
  \bibnamefont{Doneva}}, \bibinfo{journal}{Phys.\ Rev.\ D}
  \textbf{\bibinfo{volume}{99}}, \bibinfo{pages}{084011}
  (\bibinfo{year}{2019}), \eprint{1901.06379}.

\bibitem[{\citenamefont{Doneva and
  Yazadjiev}(2020{\natexlab{a}})}]{Doneva:2019krb}
\bibinfo{author}{\bibfnamefont{D.~D.} \bibnamefont{Doneva}} \bibnamefont{and}
  \bibinfo{author}{\bibfnamefont{S.~S.} \bibnamefont{Yazadjiev}},
  \bibinfo{journal}{Phys.\ Rev.\ D} \textbf{\bibinfo{volume}{101}},
  \bibinfo{pages}{024009} (\bibinfo{year}{2020}{\natexlab{a}}),
  \eprint{1909.00473}.

\bibitem[{\citenamefont{Collodel et~al.}(2020)\citenamefont{Collodel, Doneva,
  and Yazadjiev}}]{Collodel:2019uns}
\bibinfo{author}{\bibfnamefont{L.~G.} \bibnamefont{Collodel}},
  \bibinfo{author}{\bibfnamefont{D.~D.} \bibnamefont{Doneva}},
  \bibnamefont{and} \bibinfo{author}{\bibfnamefont{S.~S.}
  \bibnamefont{Yazadjiev}}, \bibinfo{journal}{Phys.\ Rev.\ D}
  \textbf{\bibinfo{volume}{101}}, \bibinfo{pages}{044021}
  (\bibinfo{year}{2020}), \eprint{1912.02498}.

\bibitem[{\citenamefont{Doneva and
  Yazadjiev}(2020{\natexlab{b}})}]{Doneva:2019ltb}
\bibinfo{author}{\bibfnamefont{D.~D.} \bibnamefont{Doneva}} \bibnamefont{and}
  \bibinfo{author}{\bibfnamefont{S.~S.} \bibnamefont{Yazadjiev}},
  \bibinfo{journal}{Phys.\ Rev.\ D} \textbf{\bibinfo{volume}{101}},
  \bibinfo{pages}{064072} (\bibinfo{year}{2020}{\natexlab{b}}),
  \eprint{1911.06908}.

\bibitem[{\citenamefont{Doneva and
  Yazadjiev}(2020{\natexlab{c}})}]{PhysRevD.101.104010}
\bibinfo{author}{\bibfnamefont{D.~D.} \bibnamefont{Doneva}} \bibnamefont{and}
  \bibinfo{author}{\bibfnamefont{S.~S.} \bibnamefont{Yazadjiev}},
  \bibinfo{journal}{Phys. Rev. D} \textbf{\bibinfo{volume}{101}},
  \bibinfo{pages}{104010} (\bibinfo{year}{2020}{\natexlab{c}}),
  \eprint{2004.03956},
  \urlprefix\url{https://link.aps.org/doi/10.1103/PhysRevD.101.104010}.

\bibitem[{\citenamefont{Pani et~al.}(2011)\citenamefont{Pani, Berti, Cardoso,
  and Read}}]{Pani:2011xm}
\bibinfo{author}{\bibfnamefont{P.}~\bibnamefont{Pani}},
  \bibinfo{author}{\bibfnamefont{E.}~\bibnamefont{Berti}},
  \bibinfo{author}{\bibfnamefont{V.}~\bibnamefont{Cardoso}}, \bibnamefont{and}
  \bibinfo{author}{\bibfnamefont{J.}~\bibnamefont{Read}},
  \bibinfo{journal}{Phys.\ Rev.\ D} \textbf{\bibinfo{volume}{84}},
  \bibinfo{pages}{104035} (\bibinfo{year}{2011}), \eprint{1109.0928}.

\bibitem[{\citenamefont{{Sotani}}(2014)}]{Sotani2014}
\bibinfo{author}{\bibfnamefont{H.}~\bibnamefont{{Sotani}}},
  \bibinfo{journal}{\prd} \textbf{\bibinfo{volume}{89}}, \bibinfo{eid}{064031}
  (\bibinfo{year}{2014}), \eprint{1402.5699}.

\bibitem[{\citenamefont{Mendes and Ortiz}(2018)}]{Mendes:2018qwo}
\bibinfo{author}{\bibfnamefont{R.~F.} \bibnamefont{Mendes}} \bibnamefont{and}
  \bibinfo{author}{\bibfnamefont{N.}~\bibnamefont{Ortiz}},
  \bibinfo{journal}{Phys.\ Rev.\ Lett.} \textbf{\bibinfo{volume}{120}},
  \bibinfo{pages}{201104} (\bibinfo{year}{2018}), \eprint{1802.07847}.

\bibitem[{\citenamefont{Sotani and Kokkotas}(2004)}]{Sotani04}
\bibinfo{author}{\bibfnamefont{H.}~\bibnamefont{Sotani}} \bibnamefont{and}
  \bibinfo{author}{\bibfnamefont{K.~D.} \bibnamefont{Kokkotas}},
  \bibinfo{journal}{\prd} \textbf{\bibinfo{volume}{70}},
  \bibinfo{pages}{084026} (\bibinfo{year}{2004}).

\bibitem[{\citenamefont{{Staykov} et~al.}(2015)\citenamefont{{Staykov},
  {Doneva}, {Yazadjiev}, and {Kokkotas}}}]{Staykov2015}
\bibinfo{author}{\bibfnamefont{K.~V.} \bibnamefont{{Staykov}}},
  \bibinfo{author}{\bibfnamefont{D.~D.} \bibnamefont{{Doneva}}},
  \bibinfo{author}{\bibfnamefont{S.~S.} \bibnamefont{{Yazadjiev}}},
  \bibnamefont{and} \bibinfo{author}{\bibfnamefont{K.~D.}
  \bibnamefont{{Kokkotas}}}, \bibinfo{journal}{\prd}
  \textbf{\bibinfo{volume}{92}}, \bibinfo{eid}{043009} (\bibinfo{year}{2015}),
  \eprint{1503.04711}.

\bibitem[{\citenamefont{{Sotani} and {Kokkotas}}(2005)}]{Sotani2005}
\bibinfo{author}{\bibfnamefont{H.}~\bibnamefont{{Sotani}}} \bibnamefont{and}
  \bibinfo{author}{\bibfnamefont{K.~D.} \bibnamefont{{Kokkotas}}},
  \bibinfo{journal}{\prd} \textbf{\bibinfo{volume}{71}}, \bibinfo{eid}{124038}
  (\bibinfo{year}{2005}).

\bibitem[{\citenamefont{Blazquez-Salcedo and
  Eickhoff}(2018)}]{Blazquez-Salcedo:2018tyn}
\bibinfo{author}{\bibfnamefont{J.~L.} \bibnamefont{Blazquez-Salcedo}}
  \bibnamefont{and} \bibinfo{author}{\bibfnamefont{K.}~\bibnamefont{Eickhoff}},
  \bibinfo{journal}{Phys.\ Rev.\ D} \textbf{\bibinfo{volume}{97}},
  \bibinfo{pages}{104002} (\bibinfo{year}{2018}), \eprint{1803.01655}.

\bibitem[{\citenamefont{Blazquez-Salcedo
  et~al.}(2016)\citenamefont{Blazquez-Salcedo, Gonzalez-Romero, Kunz, Mojica,
  and Navarro-Lerida}}]{Blazquez-Salcedo:2015ets}
\bibinfo{author}{\bibfnamefont{J.~L.} \bibnamefont{Blazquez-Salcedo}},
  \bibinfo{author}{\bibfnamefont{L.~M.} \bibnamefont{Gonzalez-Romero}},
  \bibinfo{author}{\bibfnamefont{J.}~\bibnamefont{Kunz}},
  \bibinfo{author}{\bibfnamefont{S.}~\bibnamefont{Mojica}}, \bibnamefont{and}
  \bibinfo{author}{\bibfnamefont{F.}~\bibnamefont{Navarro-Lerida}},
  \bibinfo{journal}{Phys.\ Rev.\ D} \textbf{\bibinfo{volume}{93}},
  \bibinfo{pages}{024052} (\bibinfo{year}{2016}), \eprint{1511.03960}.

\bibitem[{\citenamefont{Blazquez-Salcedo
  et~al.}(2020)\citenamefont{Blazquez-Salcedo, Khoo, and
  Kunz}}]{Blazquez-Salcedo:2020ibb}
\bibinfo{author}{\bibfnamefont{J.~L.} \bibnamefont{Blazquez-Salcedo}},
  \bibinfo{author}{\bibfnamefont{F.~S.} \bibnamefont{Khoo}}, \bibnamefont{and}
  \bibinfo{author}{\bibfnamefont{J.}~\bibnamefont{Kunz}}
  (\bibinfo{year}{2020}), \eprint{2001.09117}.

\bibitem[{\citenamefont{Altaha~Motahar
  et~al.}(2019)\citenamefont{Altaha~Motahar, Blazquez-Salcedo, Doneva, Kunz,
  and Yazadjiev}}]{AltahaMotahar:2019ekm}
\bibinfo{author}{\bibfnamefont{Z.}~\bibnamefont{Altaha~Motahar}},
  \bibinfo{author}{\bibfnamefont{J.~L.} \bibnamefont{Blazquez-Salcedo}},
  \bibinfo{author}{\bibfnamefont{D.~D.} \bibnamefont{Doneva}},
  \bibinfo{author}{\bibfnamefont{J.}~\bibnamefont{Kunz}}, \bibnamefont{and}
  \bibinfo{author}{\bibfnamefont{S.~S.} \bibnamefont{Yazadjiev}},
  \bibinfo{journal}{Phys.\ Rev.\ D} \textbf{\bibinfo{volume}{99}},
  \bibinfo{pages}{104006} (\bibinfo{year}{2019}), \eprint{1902.01277}.

\bibitem[{\citenamefont{Altaha~Motahar
  et~al.}(2018)\citenamefont{Altaha~Motahar, Blazquez-Salcedo, Kleihaus, and
  Kunz}}]{AltahaMotahar:2018djk}
\bibinfo{author}{\bibfnamefont{Z.}~\bibnamefont{Altaha~Motahar}},
  \bibinfo{author}{\bibfnamefont{J.~L.} \bibnamefont{Blazquez-Salcedo}},
  \bibinfo{author}{\bibfnamefont{B.}~\bibnamefont{Kleihaus}}, \bibnamefont{and}
  \bibinfo{author}{\bibfnamefont{J.}~\bibnamefont{Kunz}},
  \bibinfo{journal}{Phys.\ Rev.\ D} \textbf{\bibinfo{volume}{98}},
  \bibinfo{pages}{044032} (\bibinfo{year}{2018}), \eprint{1807.02598}.

\bibitem[{\citenamefont{Blazquez-Salcedo
  et~al.}(2019)\citenamefont{Blazquez-Salcedo, Altaha~Motahar, Doneva, Khoo,
  Kunz, Mojica, Staykov, and Yazadjiev}}]{Blazquez-Salcedo:2018pxo}
\bibinfo{author}{\bibfnamefont{J.~L.} \bibnamefont{Blazquez-Salcedo}},
  \bibinfo{author}{\bibfnamefont{Z.}~\bibnamefont{Altaha~Motahar}},
  \bibinfo{author}{\bibfnamefont{D.~D.} \bibnamefont{Doneva}},
  \bibinfo{author}{\bibfnamefont{F.~S.} \bibnamefont{Khoo}},
  \bibinfo{author}{\bibfnamefont{J.}~\bibnamefont{Kunz}},
  \bibinfo{author}{\bibfnamefont{S.}~\bibnamefont{Mojica}},
  \bibinfo{author}{\bibfnamefont{K.~V.} \bibnamefont{Staykov}},
  \bibnamefont{and} \bibinfo{author}{\bibfnamefont{S.~S.}
  \bibnamefont{Yazadjiev}}, \bibinfo{journal}{Eur.\ Phys.\ J.\ Plus}
  \textbf{\bibinfo{volume}{134}}, \bibinfo{pages}{46} (\bibinfo{year}{2019}),
  \eprint{1810.09432}.

\bibitem[{\citenamefont{{Akmal} et~al.}(1998)\citenamefont{{Akmal},
  {Pandharipande}, and {Ravenhall}}}]{AkmalPR}
\bibinfo{author}{\bibfnamefont{A.}~\bibnamefont{{Akmal}}},
  \bibinfo{author}{\bibfnamefont{V.~R.} \bibnamefont{{Pandharipande}}},
  \bibnamefont{and} \bibinfo{author}{\bibfnamefont{D.~G.}
  \bibnamefont{{Ravenhall}}}, \bibinfo{journal}{\prc}
  \textbf{\bibinfo{volume}{58}}, \bibinfo{pages}{1804} (\bibinfo{year}{1998}).

\bibitem[{\citenamefont{{Read} et~al.}(2009)\citenamefont{{Read}, {Lackey},
  {Owen}, and {Friedman}}}]{Read2009}
\bibinfo{author}{\bibfnamefont{J.~S.} \bibnamefont{{Read}}},
  \bibinfo{author}{\bibfnamefont{B.~D.} \bibnamefont{{Lackey}}},
  \bibinfo{author}{\bibfnamefont{B.~J.} \bibnamefont{{Owen}}},
  \bibnamefont{and} \bibinfo{author}{\bibfnamefont{J.~L.}
  \bibnamefont{{Friedman}}}, \bibinfo{journal}{\prd}
  \textbf{\bibinfo{volume}{79}}, \bibinfo{eid}{124032} (\bibinfo{year}{2009}),
  \eprint{0812.2163}.

\bibitem[{\citenamefont{J.Ruoff}(2000)}]{RuoffPhD}
\bibinfo{author}{\bibnamefont{J.Ruoff}}, Ph.D. thesis,
  \bibinfo{school}{University ot Tuebingen} (\bibinfo{year}{2000}).

\bibitem[{\citenamefont{Kokkotas and Ruoff}(2001)}]{Kokkotas:2000up}
\bibinfo{author}{\bibfnamefont{K.}~\bibnamefont{Kokkotas}} \bibnamefont{and}
  \bibinfo{author}{\bibfnamefont{J.}~\bibnamefont{Ruoff}},
  \bibinfo{journal}{Astron.\ Astrophys.} \textbf{\bibinfo{volume}{366}},
  \bibinfo{pages}{565} (\bibinfo{year}{2001}), \eprint{gr-qc/0011093}.

\bibitem[{\citenamefont{{Kokkotas} and {Schutz}}(1986)}]{1986GReGr..18..913K}
\bibinfo{author}{\bibfnamefont{K.~D.} \bibnamefont{{Kokkotas}}}
  \bibnamefont{and} \bibinfo{author}{\bibfnamefont{B.~F.}
  \bibnamefont{{Schutz}}}, \bibinfo{journal}{General Relativity and
  Gravitation} \textbf{\bibinfo{volume}{18}}, \bibinfo{pages}{913}
  (\bibinfo{year}{1986}).

\bibitem[{\citenamefont{{Kokkotas} and {Schutz}}(1992)}]{1992MNRAS.255..119K}
\bibinfo{author}{\bibfnamefont{K.~D.} \bibnamefont{{Kokkotas}}}
  \bibnamefont{and} \bibinfo{author}{\bibfnamefont{B.~F.}
  \bibnamefont{{Schutz}}}, \bibinfo{journal}{\mnras}
  \textbf{\bibinfo{volume}{255}}, \bibinfo{pages}{119} (\bibinfo{year}{1992}).

\bibitem[{\citenamefont{Gerosa et~al.}(2016)\citenamefont{Gerosa, Sperhake, and
  Ott}}]{Gerosa:2016fri}
\bibinfo{author}{\bibfnamefont{D.}~\bibnamefont{Gerosa}},
  \bibinfo{author}{\bibfnamefont{U.}~\bibnamefont{Sperhake}}, \bibnamefont{and}
  \bibinfo{author}{\bibfnamefont{C.~D.} \bibnamefont{Ott}},
  \bibinfo{journal}{Class. Quant. Grav.} \textbf{\bibinfo{volume}{33}},
  \bibinfo{pages}{135002} (\bibinfo{year}{2016}), \eprint{1602.06952}.

\end{thebibliography}

\end{document}